\documentclass[
    aps,
    prd,
    reprint,
    superscriptaddress,
    amsmath,
    amssymb,
    floatfix,
    nofootinbib
]{revtex4-2}
\usepackage{diagbox}

\usepackage{graphicx}
\usepackage[caption=false]{subfig}

\usepackage{bm}
\usepackage{mathtools}

\usepackage[dvipsnames]{xcolor}

\usepackage[normalem]{ulem}

\definecolor{greenW}{rgb}{0.0,0.55,0.1}

\usepackage{hyperref}
\hypersetup{
    colorlinks=true,
    linkcolor=blue,
    citecolor=red,
    urlcolor=greenW
}

\begin{document}


\title{Probing generalized equation of state for dark energy with latest cosmological probes}

\author{Dadian Cheng}
\affiliation{
School of Physics and Astronomy,
Beijing Normal University,
Beijing 100875, China
}

\author{Shuo Cao}
\email{caoshuo@bnu.edu.cn}
\affiliation{
School of Physics and Astronomy,
Beijing Normal University,
Beijing 100875, China
}
\affiliation{
Institute for Frontiers in Astronomy and Astrophysics,
Beijing Normal University,
Beijing 102206, China
}

\author{Yalong Nan}
\affiliation{
School of Physics and Astronomy,
Beijing Normal University,
Beijing 100875, China
}
\affiliation{
Institute for Frontiers in Astronomy and Astrophysics,
Beijing Normal University,
Beijing 102206, China
}

\author{Xinyue Jiang}
\affiliation{
School of Physics and Astronomy,
Beijing Normal University,
Beijing 100875, China
}
\affiliation{
Institute for Frontiers in Astronomy and Astrophysics,
Beijing Normal University,
Beijing 102206, China
}

\author{Sihan Li}
\affiliation{
School of Physics and Astronomy,
Beijing Normal University,
Beijing 100875, China
}
\affiliation{
Institute for Frontiers in Astronomy and Astrophysics,
Beijing Normal University,
Beijing 102206, China
}

\author{Mingming Wang}
\affiliation{
School of Physics and Astronomy,
Beijing Normal University,
Beijing 100875, China
}
\affiliation{
Institute for Frontiers in Astronomy and Astrophysics,
Beijing Normal University,
Beijing 102206, China
}

\author{Shijie Wang}
\affiliation{
School of Physics and Astronomy,
Beijing Normal University,
Beijing 100875, China
}
\affiliation{
Institute for Frontiers in Astronomy and Astrophysics,
Beijing Normal University,
Beijing 102206, China
}

\begin{abstract}

Baryon acoustic oscillation (BAO) measurements from DESI DR2 have strengthened indications of a time-evolving dark energy. In this paper, we investigate observational constraints on the generalized equation of state (GEoS) for dark energy, $w(a) = w_0 - w_\beta (a^{\beta}- 1)/\beta$, with latest data from UV and X-ray observations of high-redshift quasars, DESI DR2 BAO, Cosmic Microwave Background (CMB) data (including lensing) from Planck and the Atacama Cosmology Telescope (ACT), and Pantheon+ Type Ia Supernovae (SNe Ia). The full Quasar+BAO+CMB+SNe Ia data combination yields stringent constraints on the EoS parameters of $w_0=-0.840^{+0.057}_{-0.084}$,
$w_\beta=-0.90^{+0.88}_{-0.23}$, and $\beta=2.2^{+2.4}_{-5.2}$. Specially, the $\beta$ parameter, which describe a broad range of dark energy evolution, is in perfect agreement with the Chevallier--Polarski--Linder model ($\beta=1$) -- the most widely used two-parameter parametrizations of dynamical dark energy. We also compare GEoS with other dynamical dark-energy parametrizations including CPL, Barboza--Alcaniz (BA) and Jassal--Bagla--Padmanabhan (JBP), and report three key findings. Firstly, we find compelling evidence for dynamical dark energy: all four EoS parametrizations yield lower information criterion (IC) values than $\Lambda$CDM for the full data combination, indicating a preference for EoS parametrizations under the Akaike information criterion (AIC) and the variance-based deviance information criterion (DIC). Secondly, although the unified phenomenological framework GEoS performs better than JBP, the current data do not strongly discriminate among different EoS parametrizations. Thirdly, the EoS trajectories of GEoS, CPL, and BA exhibit qualitatively similar transitions from the phantom regime at higher redshifts to the quintessence regime at late times, while the JBP parametrization exhibits a nonmonotonic evolution.

\end{abstract}

\maketitle

\section{Introduction}

Over the past decades, the $\Lambda$CDM model~\cite{1984ApJ...284..439P} has become the standard framework of modern cosmology. The late-time accelerated expansion of the Universe was first established through observations of Type Ia supernovae (SNe Ia)~\cite{1998AJ....116.1009R,1999ApJ...517..565P}. Subsequent observations of large-scale structure~\cite{2005ApJ...633..560E} and the CMB~\cite{2003ApJS..148..175S} have provided increasingly precise constraints on its cosmological parameters. Within this framework, dark energy is described by a cosmological
constant $\Lambda$, corresponding to an EoS $w=-1$. Although a wide range of independent cosmological observations~\cite{2020A&A...641A...6P,2021PhRvD.103h3533A,2022ApJ...938..110B} are broadly consistent with the $\Lambda$CDM model, several theoretical and observational challenges remain. From a theoretical perspective, the cosmological constant is associated with the long-standing cosmological constant problem~\cite{1989RvMP...61....1W,2000IJMPD...9..373S} and the cosmic coincidence problem~\cite{1999PhRvL..82..896Z}. On the observational side, discrepancies have emerged between some early- and late-Universe determinations of cosmological parameters. In particular, the value of the Hubble constant ($H_0$) inferred from Planck CMB observations~\cite{2020A&A...641A...6P} within the $\Lambda$CDM framework differs significantly from late-time distance-ladder measurements, such as those obtained by the SH0ES collaboration using Cepheid-calibrated SNe Ia~\cite{2022ApJ...934L...7R}. This discrepancy, commonly referred to as the $H_0$ tension, has motivated extensive investigations of both possible systematic effects and extensions beyond the standard cosmological model~\cite{2019NatAs...3..891V,2021APh...13102605D,2022PhR...984....1S,2021A&ARv..29....9S,perivolaropoulos2024hubbletensiondistanceladder,2026MNRAS.548ag584L}. A further discrepancy concerns the amplitude of late-time matter clustering, commonly quantified through $S_8$, with some weak-lensing and large-scale-structure measurements preferring lower values than those inferred from CMB observations within $\Lambda$CDM~\cite{2021APh...13102604D}. These theoretical and observational issues have motivated extensive studies of alternative cosmological scenarios~\cite{Koyama_2016,Bahamonde_2023,Bili__2002,Padmanabhan_2002}.

More recently, the DESI DR2 has provided high-precision BAO measurements over a broad redshift range, $0.1\lesssim z\lesssim4.2$. When combined with CMB observations and different SNe Ia compilations, the DESI DR2 BAO measurements strengthen the indications for a time-varying dark energy EoS. In particular, within the CPL parametrization~\cite{CHEVALLIER_2001,Linder_2003}, the preference over the flat $\Lambda$CDM model reaches approximately $2.8\sigma$--$4.2\sigma$, depending on the SNe Ia dataset included in the joint analysis~\cite{Abdul_Karim_2025}. These results have stimulated renewed investigations of the possible dynamical nature of dark energy and its underlying physical mechanisms~\cite{2026PDU....5202254L,özülker2025darkenergycrossesline,Cheng_2025,Fazzari_2025}. Related studies have also explored whether modifications to other components of the cosmological model, including the dark matter sector, could contribute to the observed phenomenology~\cite{Pan_2026,Silva_2025,Guedezounme_2026,Roy_Choudhury_2024}.
Motivated by these developments, we investigate dynamical dark energy(DDE) parametrizations: the GEoS parametrization~\cite{Barboza_2009}, the CPL parametrization~\cite{CHEVALLIER_2001,Linder_2003}, the BA parametrization~\cite{Barboza_2008}, and the JBP parametrization~\cite{Jassal_2005}. In particular, the GEoS parametrization is characterized by three free parameters, $w_0$, $w_\beta$, and $\beta$. This parametrization encompasses several commonly studied forms of dynamical dark energy as limiting cases: it reduces to the CPL parametrization for $\beta=1$, approaches the logarithmic parametrization~\cite{Efstathiou_1999} in the limit $\beta\to0$, and recovers the linear parametrization~\cite{Cooray_1999} for $\beta=-1$. The additional parameter $\beta$ therefore provides greater flexibility in describing the redshift evolution of the dark energy EoS. By comparing the GEoS parametrization with the CPL, BA, and JBP parametrizations, we aim to investigate whether current observations support this additional freedom in the evolution of dark energy.

Extending cosmological distance measurements beyond the redshift range
covered by SNe Ia samples can provide additional information
for constraining the redshift evolution of the dark-energy EoS.
A key ingredient in quasar cosmology is the empirical nonlinear
correlation between their rest-frame UV and X-ray luminosities~\cite{Lusso_2016,Lusso_2020,Bisogni_2021}.
After carefully accounting for observational selection effects,
intrinsic dispersion, and possible model dependence, this relation
can be used to construct a quasar Hubble diagram, allowing quasars
to serve as complementary standardized probes for cosmological studies~\cite{
risaliti2018cosmologicalconstraintshubblediagram,2020ApJ...899...71L,2020MNRAS.496..708L,2020ApJ...892..103Z,Bargiacchi_2021}.
Their broad redshift coverage extends the distance--redshift relation
well beyond the range sampled by conventional SNe Ia,
providing additional constraints on the redshift evolution of the
dark-energy EoS. Quasar observations have therefore increasingly been combined with other cosmological probes to constrain dark energy
models~\cite{2021SCPMA..6459511Z,Bargiacchi_2023,Dainotti_2023,
benetti2025quasarcosmologyiijoint}. In this work, we use multiple measurements of quasars in combination with several complementary cosmological probes to constrain four DDE parametrizations. Specifically, we incorporate BAO measurements from DESI DR2, CMB observations from Planck and ACT, and SNe Ia data from the Pantheon+ compilation. The combination of these probes spans a broad redshift range and helps reduce parameter degeneracies, allowing us to examine the redshift evolution of the dark-energy EoS and to compare the performance of the different parametrizations. The remainder of this paper is organized as follows. In Section~\ref{sec:dde_models}, we present the theoretical framework of the cosmological field equations and introduce the DDE models adopted in this work. In Section~\ref{sec:data_methodology}, we describe the observational datasets and elaborate on the statistical methodology employed to constrain these models. In Section~\ref{sec:results}, we analyze the cosmological constraints derived for the considered DDE models. Finally, Section~\ref{sec:conclusions} summarizes the conclusions.

\section{Dynamical Dark Energy Models}
\label{sec:dde_models}
According to the cosmological principle, the Universe is homogeneous
and isotropic on sufficiently large scales and can therefore be
described by the Friedmann--Lemaître--Robertson--Walker (FLRW)
geometry. Current cosmological observations place tight constraints
on the spatial-curvature parameter $\Omega_k$, with the results being
consistent with a spatially flat Universe~\cite{CAO2019100274,2020ApJ...901..129L,2025ApJ...981L..24L}.
Motivated by these observational constraints, we assume a spatially
flat Universe throughout this work, for which the FLRW metric takes
the form of 
\begin{equation}
    ds^2 = -dt^2 + a^2(t)(dx^2 + dy^2 + dz^2) \,,
\end{equation}
Here $a(t)$ denotes the cosmic scale factor, $t$ is the cosmic time, and $(x,y,z)$ represent the spatial coordinates. Within the framework of general relativity, we assume that no non-gravitational interactions exist between any two cosmic fluid components. The evolution of the Universe can then be determined by solving the Einstein field equations.

We consider a universe composed of radiation, pressureless matter
(baryons and cold dark matter), neutrinos, and dark energy. For a
spatially flat Universe, the Friedmann equation is given by
\begin{equation}
    H^2(a)=\frac{8\pi G}{3}
    \left(
    \rho_r+\rho_m+\rho_\nu+\rho_{\rm DE}
    \right),
    \label{eq:hubble}
\end{equation}
where $\rho_r$, $\rho_m$, $\rho_\nu$, and $\rho_{\rm DE}$ denote
the energy densities of radiation, pressureless matter, neutrinos,
and dark energy, respectively. Assuming that the individual components are separately conserved, each component satisfies the continuity equation
\begin{equation}
    \dot{\rho}_i+3H(1+w_i)\rho_i=0,
\end{equation}
where $w_i=p_i/\rho_i$ denotes the EoS parameter of the $i$th
component, with $\rho_i$ and $p_i$ being its energy density and
pressure, respectively. We normalize the present-day scale factor
to $a_0=1$. For radiation, with $w_r=1/3$, and pressureless matter,
with $w_m=0$, the corresponding energy densities evolve as
\begin{equation}
    \rho_r(a)=\rho_{r,0}a^{-4},
    \qquad
    \rho_m(a)=\rho_{m,0}a^{-3},
\end{equation}
where $\rho_{r,0}$ and $\rho_{m,0}$ denote their present-day energy
densities. For a general time-dependent dark-energy EoS $w(a)$, the continuity
equation gives
\begin{equation}
    \rho_{\rm DE}(a)
    =
    \rho_{{\rm DE},0}
    \exp\left[
    -3\int_1^a
    \frac{1+w(a')}{a'}\,da'
    \right],
    \label{eq:rho_de}
\end{equation}
where $\rho_{{\rm DE},0}$ denotes the present-day dark-energy density.
For a cosmological constant, $w=-1$ and $\rho_{\rm DE}$ remains
constant, whereas the DDE models considered in this work allow
$w$ to evolve with the scale factor. The total neutrino mass is fixed to
$\sum m_\nu=0.06\,{\rm eV}$, while the effective number of
relativistic species is fixed to $N_{\rm eff}=3.044$.

In addition to the background evolution, the calculation of
cosmological observables requires the evolution of linear
perturbations. The dark-energy EoS affects both the background
expansion and the evolution of cosmological perturbations. To describe scalar perturbations, we work in the synchronous
gauge~\cite{Ma_1995}, for which the perturbed line element is
\begin{equation}
    ds^2=a^2(\tau)
    \left[
    -d\tau^2+
    \left(\delta_{ij}+h_{ij}\right)dx^i dx^j
    \right],
\end{equation}
where $\tau$ denotes conformal time, $\delta_{ij}$ is the background
spatial metric for a spatially flat Universe, and $h_{ij}$ denotes
the metric perturbation. For the $i$th fluid component, we define
the density contrast as $\delta_i=\delta\rho_i/\rho_i$, while
$\theta_i$ denotes the divergence of its peculiar velocity. In
Fourier space, their linear evolution equations can be written as
~\cite{Ma_1995}
\begin{equation}
\begin{split}
\delta_i'={}&
-(1+w_i)
\left(\theta_i+\frac{h'}{2}\right)
-3\mathcal{H}(c_{s,i}^2-w_i)\delta_i
\\
&-9\mathcal{H}^2
(1+w_i)
(c_{s,i}^2-c_{a,i}^2)
\frac{\theta_i}{\kappa^2},
\end{split}
\label{eq:delta}
\end{equation}
and
\begin{equation}
\theta_i'
=
-\mathcal{H}(1-3c_{s,i}^2)\theta_i
+\frac{c_{s,i}^2}{1+w_i}\kappa^2\delta_i
-\kappa^2\sigma_i ,
\label{eq:theta}
\end{equation}
where a prime denotes a derivative with respect to conformal time,
$\mathcal{H}=a'/a$ is the conformal Hubble parameter, $h$ is the
trace metric perturbation in the synchronous gauge, $\kappa$ is the
comoving wavenumber, and $\sigma_i$ denotes the anisotropic stress.
The rest-frame sound speed and the adiabatic sound speed are defined
as
\begin{equation}
c_{s,i}^2
\equiv
\left.
\frac{\delta p_i}{\delta\rho_i}
\right|_{\rm rest},
\qquad
c_{a,i}^2
\equiv
\frac{p_i'}{\rho_i'}
=
w_i-\frac{w_i'}
{3\mathcal{H}(1+w_i)}.
\end{equation}
For dark energy, we assume vanishing anisotropic stress,
$\sigma_{\rm DE}=0$, and fix its rest-frame sound speed to
$c_{s,\rm DE}^2=1$.

Since the dark-energy EoS affects both the background expansion and
the evolution of cosmological perturbations, we consider four
phenomenological DDE parametrizations for $w(a)$. We consider the GEoS parametrization featuring three free parameters~\cite{Barboza_2009}
\begin{equation}
    w(a) = w_0 - w_\beta \left[ \frac{a^\beta - 1}{\beta} \right] \,,
    \label{eq:general}
\end{equation}
where $w_0$ denotes the present-day value of the DE EoS, $w_\beta$ characterizes the dynamical properties of DE, and $\beta$ governs a wide variety of DE evolutionary behaviors.  Substituting Eq.~(\ref{eq:general}) into Eq.~(\ref{eq:rho_de}), the corresponding energy density evolution takes the exact analytical form
\begin{equation}
    \frac{\rho_{DE}(a)}{\rho_{DE,0}} = a^{-3\left(1 + w_0 + \frac{w_\beta}{\beta}\right)} \exp \left[ \frac{3 w_\beta}{\beta^2} (a^\beta - 1) \right] \,.
\end{equation}
The most compelling feature of this generalized parametrization is its ability to recover multiple existing DE models naturally depending on the $\beta$ parameter:
\begin{itemize}
    \item For $\beta = 1$, it reduces exactly to the CPL parametrization: $w(a) = w_0 + w_a(1-a)$ (with $w_a = w_\beta$).
    \item For $\beta = -1$, it recovers the linear parametrization~\cite{Cooray_1999}: $w(a) = w_0 + w_\beta(a^{-1} - 1) = w_0 + w_\beta z$.
    \item In the limit where $\beta \to 0$, it converges to the logarithmic parametrization~\cite{Efstathiou_1999}: $w(a) = w_0 - w_\beta \ln a$.
\end{itemize}
The GEoS parametrization therefore provides a unified phenomenological
framework that includes several commonly studied DDE parametrizations
as special cases. Constraints on $\beta$ can then be used to examine
which functional forms of the dark-energy EoS are consistent with
the observational data \cite{2014PhRvD..90h3006C,2017EPJC...77..891M}.

In this work, we also conduct a systematic comparison of other three representative DDE frameworks using the latest observational data. Firstly, the Chevallier--Polarski--Linder model (CPL) parametrization~\cite{CHEVALLIER_2001,Linder_2003} is one of the most widely used two-parameter parametrizations of DDE, which has been extensively employed in cosmological analysis~\cite{2012JCAP...03..016C,2015ApJ...806..185C,Abdul_Karim_2025}. In this parametrization, the dark-energy EoS is written as
\begin{equation}
    w(a)=w_0+w_a(1-a),
    \label{eq:cpl}
\end{equation}
where $w_0$ denotes the present-day value of the EoS, while $w_a$
characterizes its evolution with the scale factor, with
$w_a=-dw/da$. In terms of redshift, the CPL parametrization can be
written as $w(z)=w_0+w_a z/(1+z)$. At low redshift, it approaches
the linear form $w(z)\simeq w_0+w_a z$, while at high redshift it
remains finite, approaching $w_0+w_a$ as $z\to\infty$. This behavior
avoids the high-redshift divergence of a simple linear 
parametrization. Secondly, we consider the Barboza--Alcaniz (BA) parametrization ~\cite{Barboza_2008}, the functional form of which is given by
\begin{equation}
w(a) = w_0 + w_a \frac{1-a}{a^2+(1-a)^2}.
\label{eq:ba}
\end{equation}
Equivalently, in terms of redshift, it can be written as
$w(z)=w_0+w_a\frac{z(1+z)}{1+z^2}$. Similar to the CPL parametrization, the BA parametrization reduces to the linear form $w(z)\simeq w_0+w_a z$ at low redshift and approaches the finite asymptotic value $w_0+w_a$ as $z\to\infty$. Despite sharing the same low- and high-redshift limits, the CPL and BA parametrizations exhibit different redshift dependence at intermediate redshifts. Moreover, unlike the CPL parametrization, the BA form remains finite in the far-future limit $z\to-1$. Thirdly, to capture potential dynamical evolution, the Jassal--Bagla--Padmanabhan (JBP) parametrization 
~\cite{Jassal_2005} introduces the EoS redshift dependence through
\begin{equation}
    w(a)=w_0+w_a a(1-a) \,.
    \label{eq:jbp}
\end{equation}
Equivalently, in terms of redshift, it can be written as
$w(z)=w_0+w_a z/(1+z)^2$. In contrast to the CPL and BA
parametrizations, the dynamical contribution in the JBP
parametrization vanishes both at the present epoch and in the
high-redshift limit, such that $w(a=1)=w_0$ and
$\lim\limits_{a\to 0} w(a)=w_0$. The deviation from $w_0$ is therefore
largest at intermediate redshifts, reaching its maximum amplitude
at $a=1/2$ ($z=1$).

\section{Observational Datasets And Methodology}
\label{sec:data_methodology}

In the following, we present a detailed introduction to the datasets employed to constrain the four DDE models proposed above.

 \textbf{UV and X-ray quasars}: The advances in cosmology over recent decades have been accompanied by searches for reliable standard probes at higher redshifts. To extend the distance--redshift relation beyond the redshift range densely sampled by SNe Ia, we employ the UV--X-ray luminosity relation of quasars. An empirical nonlinear relation exists between the UV and X-ray luminosities of quasars: $\log_{10} L_X = \gamma \log_{10} L_{UV} + \beta$, where $L_X$ and $L_{UV}$ denote the rest-frame monochromatic luminosities at $2\text{ keV}$ and $2500\text{ \AA}$, respectively~\cite{Lusso_2016,Lusso_2020,Bisogni_2021}. Note that the parameter $\beta$ used here differs from the $\beta$ parameter included in the GEoS model. To avoid ambiguity, we denote the former parameter as $\beta_q$. Following a model-independent calibration scheme presented in a previous work, we allow $\gamma$ and $\beta$ to evolve linearly with redshift, namely $\gamma(z) = \gamma_0 + \gamma_1(1 + z)$ and $\beta_q(z) = \beta_0 + \beta_1(1 + z)$~\cite{2025ApJ...983..141L}. The luminosity distance $D_L(z)$ (in units of cm) connects the intrinsic luminosity L to the observed flux F (in units of $\text{erg s}^{-1} \text{cm}^{-2} \text{Hz}^{-1}$) via the relation $F = L / 4\pi D_L^2$. The theoretically predicted logarithmic X-ray flux evaluated at redshift $z$ can be expressed as
      \begin{equation}
        \begin{aligned}
            \log_{10}(F_X^{\rm th}) &= \gamma(z) \log_{10}(F_{UV}) + 2[\gamma(z) - 1] \log_{10}[D_L(z)] \\
            & \quad + \beta_q(z) + [\gamma(z) - 1] \log_{10}(4\pi) \,.
        \end{aligned}
    \end{equation}
    From the full sample of 2421 quasars, we select subsamples with $z>0.7$ to mitigate residual contamination originating from host galaxies~\cite{Lusso_2020}. The log-likelihood function for the quasar sample is written as
    \begin{equation}
   \begin{aligned}
   \ln \mathcal{L}_{\rm quasar} = -\frac{1}{2} \sum_{i=1}^{N} & \left[ \frac{\left( \log_{10}(F_{X,i}^{\rm obs}) - \log_{10}(F_{X,i}^{\rm th}) \right)^2}{s_i^2} \right. \\
   & \left. \vphantom{\frac{\left(\log_{10}^{\rm obs}\right)^2}{s_i^2}} + \ln(s_i^2) \right] \,,
   \end{aligned}
   \end{equation}
    where the total variance $s_i^2$ is defined as $s_i^2 = \sigma_{\log(F_{X,i})}^2 + \gamma(z_i)^2 \sigma_{\log(F_{\mathrm{UV},i})}^2 + \delta^2$.
    Here, $\sigma_{\log(F_{X,i})}$ and $\sigma_{\log(F_{\mathrm{UV},i})}$ denote the observational 
    errors associated with the X-ray and UV logarithmic fluxes, respectively, while 
   the parameter $\delta$ accounts for the intrinsic dispersion of the quasar 
   scaling relation. The free parameters associated with quasars mentioned above have recently been calibrated via a model-independent approach using SNe Ia data, yielding $\gamma_0 = 0.638 \pm 0.065$, $\gamma_1 = -0.034 \pm 0.027$, $\beta_0 = 6.9 \pm 2.0$, $\beta_1 = 1.18 \pm 0.83$, and an intrinsic dispersion of $\delta = 0.2290 \pm 0.0042$~\cite{2025ApJ...983..141L}. We treat these quantities as nuisance
   parameters and impose Gaussian priors based on these calibration
   results. For simplicity, we collectively designate this quasar dataset as ``Q'' throughout the remainder of this work.

\textbf{Baryon Acoustic Oscillations}: We use the baryon acoustic oscillation (BAO) measurements from the second data release (DR2) of the Dark Energy Spectroscopic Instrument (DESI)~\cite{Abdul_Karim_2025}. These
measurements are derived from the clustering of the Bright Galaxy
Sample (BGS), luminous red galaxies (LRGs), emission line galaxies
(ELGs), and quasars (QSOs), together with the Ly$\alpha$ forest.
We adopt the all-tracer DESI DR2 BAO likelihood implemented in
Cobaya, which comprises 13 distance-ratio measurements at seven
effective redshifts spanning $0.295 \leq z_{\rm eff} \leq 2.33$.
The observables are $D_V(z)/r_d$ for BGS and
$D_M(z)/r_d$ and $D_H(z)/r_d$ for the remaining samples, where
\begin{equation}
    D_M(z) = (1+z)D_A(z), \qquad
    D_H(z) = \frac{c}{H(z)},
\end{equation}
and
\begin{equation}
    D_V(z) = \left[zD_M^2(z)D_H(z)\right]^{1/3}.
\end{equation}
Here, $D_A$ is the angular diameter distance and $r_d$ is the
comoving sound horizon at the baryon drag epoch. For each cosmological parameter set $\boldsymbol{\theta}$, the corresponding theoretical predictions for these distance measures and the sound horizon $r_d$ are calculated within the cosmological model under consideration. 
The Gaussian likelihood is constructed as
\begin{equation}
    \ln \mathcal{L}_{\rm BAO}
    = -\frac{1}{2}
    \boldsymbol{\Delta}_{\rm BAO}^{\rm T}
    \mathbf{C}_{\rm BAO}^{-1}
    \boldsymbol{\Delta}_{\rm BAO},
\end{equation}
where
\begin{equation}
    \boldsymbol{\Delta}_{\rm BAO}
    =
    \boldsymbol{\mathcal{D}}_{\rm th}(\boldsymbol{\theta})
    -
    \boldsymbol{\mathcal{D}}_{\rm obs}
\end{equation}
is the residual vector between the theoretical and observed BAO
distance ratios. Here,
$\boldsymbol{\mathcal{D}}_{\rm th}$ contains the theoretical values
of $D_V/r_d$, $D_M/r_d$, and $D_H/r_d$ at the corresponding
effective redshifts, while $\mathbf{C}_{\rm BAO}$ is the full
covariance matrix supplied with the DESI DR2 measurements.
Throughout this work, we denote this dataset by ``B''.

 \textbf{Cosmic Microwave Background and Lensing Data}: We use measurements of the Cosmic Microwave Background (CMB) temperature and polarization anisotropies from Planck, together with CMB lensing measurements from Planck PR4 and ACT DR6. Our CMB dataset, collectively denoted by ``C'' throughout this work, consists of the following components:
    \begin{itemize}
        \item[(1)] At small angular scales (high-multipoles, $\ell > 30$), the temperature, polarization, and cross-correlation power spectra ($C_\ell^{TT}$, $C_\ell^{EE}$, and $C_\ell^{TE}$) are derived using the NPIPE PR4 Planck \texttt{CamSpec} likelihood~\cite{Efstathiou_2021,Rosenberg_2022}.
        \item[(2)] At large angular scales (low-multipoles, $2 \leq \ell < 30$), the temperature anisotropy power spectrum, $C_\ell^{TT}$, is extracted using the Planck \texttt{Commander} likelihood~\cite{2020A&A...641A...5P,2020A&A...641A...6P}.
        \item[(3)] For the E-mode polarization at large scales ($2 \leq \ell < 30$), the power spectrum $C_\ell^{EE}$ is acquired via the Planck \texttt{SimAll} likelihood~\cite{2020A&A...641A...5P,2020A&A...641A...6P}.
        \item[(4)] The CMB lensing likelihood incorporates the most recent and accurate measurements by combining the NPIPE PR4 Planck CMB lensing\footnote{\url{https://github.com/carronj/planck_PR4_lensing.}}~\cite{Carron_2022} 
        and the ACT DR6 lensing likelihood\footnote{\url{https://github.com/ACTCollaboration/act_dr6_lenslike.}}~\cite{Qu_2024,Madhavacheril_2024}. 
    \end{itemize}

\textbf{Type Ia Supernovae}: We use the Pantheon+ compilation, which contains 1701 light curves corresponding to 1550 distinct Type Ia supernovae (SNe Ia) from 18 surveys~\cite{2022ApJ...938..110B}.
We use the Pantheon+ SN Ia likelihood without incorporating the SH0ES Cepheid-based absolute-distance calibration. Accordingly, $M_B$ is not fixed to an externally calibrated value; instead, the common absolute-magnitude offset is analytically marginalized over assuming a flat prior. The likelihood applies the cut $z_{\rm HD}>0.01$, retaining 1590 light-curve measurements, and
uses the full statistical and systematic covariance matrix,
$\mathbf{C}_{\rm SN}$, restricted to the selected sample. For each supernova measurement, the theoretical corrected
apparent magnitude is
\begin{equation}
    m_{B,i}^{\rm th}
    = M_B
    + 5\log_{10}\left(
        \frac{D_{L,i}(\boldsymbol{\theta})}{\mathrm{Mpc}}
      \right)
    + 25,
\end{equation}
where $M_B$ is the common absolute-magnitude offset.
Following the likelihood implementation, we calculate
\begin{equation}
    D_{L,i}(\boldsymbol{\theta})
    = (1+z_{{\rm HEL},i})(1+z_{{\rm HD},i})
      D_A(z_{{\rm HD},i};\boldsymbol{\theta}),
\end{equation}
where $z_{\rm HEL}$ and $z_{\rm HD}$ are the heliocentric and
Hubble-diagram redshifts, respectively. We analytically
marginalize over the common magnitude offset with a flat prior.
Defining the residual vector by
\begin{equation}
    \Delta_{{\rm SN},i}
    = m_{B,i}^{\rm corr}
    - 5\log_{10}\left(
        \frac{D_{L,i}(\boldsymbol{\theta})}{\mathrm{Mpc}}
      \right)
    - 25 - M_B,
\end{equation}
Since no external absolute-distance calibration is included, we analytically marginalize over $M_B$ assuming a flat prior. Up to parameter-independent normalization constants, the resulting log-likelihood is
\begin{equation}
    \ln \mathcal{L}_{\rm SN}
    = -\frac{1}{2}
      \left(A-\frac{B^2}{E}\right),
\end{equation}
with
\begin{align}
    A &=
    \boldsymbol{\Delta}_{\rm SN}^{\rm T}
    \mathbf{C}_{\rm SN}^{-1}
    \boldsymbol{\Delta}_{\rm SN}, \\
    B &=
    \boldsymbol{1}^{\rm T}
    \mathbf{C}_{\rm SN}^{-1}
    \boldsymbol{\Delta}_{\rm SN}, \\
    E &=
    \boldsymbol{1}^{\rm T}
    \mathbf{C}_{\rm SN}^{-1}
    \boldsymbol{1},
\end{align}
where $\boldsymbol{1}$ is a vector of ones.
This treatment constrains the relative luminosity-distance
relation without imposing an absolute distance calibration.
Throughout this work, we denote this dataset by ``S''.

To compute the theoretical predictions for the cosmological
observables, we implement the four DDE parametrizations considered
in this work in modified versions of the publicly available
Boltzmann code \texttt{CAMB} (Code for Anisotropies in the Microwave
Background)~\cite{Lewis_2000,Howlett_2012}. We sample the posterior
distributions of the cosmological and model parameters using the
MCMC sampler implemented in the publicly available Bayesian inference
framework \texttt{Cobaya} (Code for Bayesian Analysis)~\cite{Torrado_2021}. The theoretical framework underlying all the investigated DDE scenarios is characterized by a multi-dimensional parameter space. The baseline cosmology is described by the six standard $\Lambda$CDM parameters: the present-day dimensionless physical baryon density $\Omega_b h^2$, the present-day dimensionless physical cold dark matter density $\Omega_c h^2$, the optical depth to reionization $\tau$, the angular size of the sound horizon at recombination $\theta_{MC}$, the scalar spectral index $n_s$, and the  amplitude of the primordial scalar power spectrum $A_s$. To characterize the dark energy sector, we introduce additional degrees of freedom depending on the specific parametrization. For the CPL, BA, and JBP models, the dark energy evolution is dictated by the present-day EoS $w_0$ and its dynamical component $w_a$. Meanwhile, the GEoS model relies on $w_0$, $w_\beta$, and $\beta$. In our Bayesian inference framework, we adopt broad uniform priors for the cosmological and DDE parameters. The quasar scaling-relation parameters are treated as nuisance parameters with Gaussian priors based on the calibration results described above. The convergence of the MCMC chains is assessed using the Gelman--Rubin convergence diagnostic $R-1$~\cite{Gelman:1992}. For all models and dataset combinations, we require $R-1<0.01$ as the convergence criterion. The resulting MCMC chains are post-processed using the publicly available package \texttt{GetDist}~\cite{Lewis_2025}.

To quantify differences in goodness of fit among the cosmological models, we evaluate the data-only $\chi^2$ at the maximum-a-posteriori (MAP) point of each converged MCMC chain,
\begin{equation}
\chi^2_{\rm MAP}
\equiv
-2\ln\mathcal{L}(\boldsymbol{\Theta}_{\rm MAP}),
\end{equation}
where $\boldsymbol{\Theta}_{\rm MAP}$ denotes the parameter vector at the MAP point. We define the difference relative to the flat $\Lambda$CDM model as
\begin{equation}
\Delta\chi^2_{\rm MAP}
\equiv
\chi^2_{{\rm MAP},i}
-
\chi^2_{{\rm MAP},\Lambda{\rm CDM}},
\end{equation}
where $i$ denotes one of the DDE models considered in this work. A negative value of $\Delta\chi^2_{\rm MAP}$ therefore indicates an improvement in goodness of fit relative to $\Lambda$CDM. However, $\Delta\chi^2_{\rm MAP}$ alone does not account for differences in model complexity. We therefore complement the goodness-of-fit comparison with the AIC and DIC, which incorporate
model complexity in different ways.

We first consider the Akaike information criterion (AIC) as a
complementary measure of model performance~\cite{1100705}. The AIC
is defined as
\begin{equation}
    {\rm AIC}
    =
    -2\ln\mathcal{L}_{\max}+2k,
\end{equation}
where $\mathcal{L}_{\max}$ denotes the maximum likelihood achievable
by the model and $k$ is the number of independently varied model
parameters. The first term measures the best-fit quality, while the
second penalizes the introduction of additional free parameters.
A lower AIC therefore indicates a more favorable balance between
goodness of fit and model complexity. For a model $i$ relative to a reference model $j$, we define $\Delta{\rm AIC}_{ij}\equiv{\rm AIC}_i-{\rm AIC}_j$. According to this convention, $\Delta{\rm AIC}_{ij}<0$ indicates that model $i$ has the lower AIC, showing a preference for model $i$ over the reference model $j$. We use the magnitude of $\Delta{\rm AIC}$ only as a descriptive measure of model discrimination. Values $|\Delta{\rm AIC}|\lesssim2$ indicate weak discrimination, differences of $4$--$7$ indicate moderate
discrimination, and values greater than $10$ indicate strong
discrimination between the two models, favoring the model with the lower AIC \cite{BurnhamAnderson2002,2011PhRvD..84b3005C}.

Whereas the AIC penalizes model complexity through the nominal
number of free parameters, the deviance information criterion (DIC)~\cite{Spiegelhalter2002,gelman2013understandingpredictiveinformationcriteria} provides a posterior-based measure of effective model complexity.
For a parameter vector $\boldsymbol{\Theta}$, the deviance is defined as
\begin{equation}
D(\boldsymbol{\Theta})
\equiv
-2\ln\mathcal{L}(\boldsymbol{\Theta}),
\end{equation}
up to an additive constant that is common to the models being compared.
The posterior mean deviance is defined as
\begin{equation}
\overline{D}
\equiv
\left\langle
D(\boldsymbol{\Theta})
\right\rangle_{\rm post}.
\end{equation}
For the effective model complexity, we adopt the variance-based estimator~\cite{gelman2013understandingpredictiveinformationcriteria}
\begin{equation}
p_V
\equiv
\frac{1}{2}
{\rm Var}_{\rm post}
\left[
D(\boldsymbol{\Theta})
\right],
\end{equation}
which provides an alternative to the conventional mean-point estimator, $p_D=\overline{D}-D(\overline{\boldsymbol{\Theta}})$, and does not require evaluating the deviance at a single posterior-mean
parameter point. The DIC used throughout this work is then defined as
\begin{equation}
{\rm DIC}
=
\overline{D}
+
p_V.
\label{eq:dic}
\end{equation}
Here, $\overline{D}$ characterizes the average goodness of fit over the
posterior distribution, while $p_V$ provides a variance-based estimate
of the effective model complexity. Smaller DIC values therefore
correspond to a more favorable balance between goodness of fit and
effective model complexity. To quantify the relative performance of two models, we define the DIC difference as $\Delta{\rm DIC}_{ij}\equiv{\rm DIC}_i-{\rm DIC}_j$, where model $j$ denotes the reference model. According to this convention, $\Delta{\rm DIC}_{ij}<0$ indicates that model $i$ has a lower DIC than the reference model, showing a preference for model $i$. The magnitude $|\Delta{\rm DIC}_{ij}|$ characterizes the degree of discrimination between the two models \citep{2021MNRAS.505.2111L,2022EPJC...82..582Z}. Since there is no uniquely defined threshold for interpreting DIC differences, we use their magnitude only as a heuristic guide for model comparison. Following commonly adopted criteria in cosmological model comparisons~\cite{Joudaki_2016,Joudaki_2017}, we take
$|\Delta{\rm DIC}|=5$ and $|\Delta{\rm DIC}|=10$ as
indicative of moderate and strong differences, respectively, in
favor of the model with the lower DIC. Smaller values indicate
correspondingly weaker discrimination between the competing models.

\begin{table*}[t]
    \centering
    \small
    \begin{tabular*}{\linewidth}{@{\extracolsep{\fill}} l c c c c c c}
        \hline\hline
         & \multicolumn{3}{c}{\textbf{GEoS}} & \multicolumn{3}{c}{\textbf{CPL}} \\
        \cline{2-4} \cline{5-7}
        \textbf{Parameters} & \textbf{QB} & \textbf{QBC} & \textbf{QBCS} & \textbf{QB} & \textbf{QBC} & \textbf{QBCS} \\
        \hline
        $100\theta_{\rm MC}$
        & --- & $1.04083\pm0.00024$ & $1.04087\pm0.00024$
        & --- & $1.04080\pm0.00024$ & $1.04086\pm0.00024$ \\
        $\Omega_b h^2$
        & $0.047^{+0.021}_{-0.010}$ & $0.02223\pm0.00013$ & $0.02225\pm0.00013$
        & $0.0475^{+0.021}_{-0.0098}$ & $0.02221\pm0.00012$ & $0.02224\pm0.00012$ \\
        $\Omega_c h^2$
        & $0.189^{+0.053}_{-0.059}$ & $0.11936\pm0.00086$ & $0.11900\pm0.00088$
        & $0.205^{+0.076}_{-0.052}$ & $0.11959\pm0.00081$ & $0.11909\pm0.00078$ \\
        $n_s$
        & --- & $0.9650\pm0.0036$ & $0.9660\pm0.0037$
        & --- & $0.9646\pm0.0035$ & $0.9657\pm0.0035$ \\
        $\tau_{\rm reio}$
        & --- & $0.0540\pm0.0072$ & $0.0561^{+0.0068}_{-0.0076}$
        & --- & $0.0527\pm0.0068$ & $0.0557\pm0.0069$ \\
        $\log(10^{10} A_s)$
        & --- & $3.041\pm0.013$ & $3.046\pm0.014$
        & --- & $3.039\pm0.013$ & $3.045\pm0.013$ \\
        \hline
        $w_0$
        & $-0.67^{+0.21}_{-0.34}$ & $-0.54^{+0.26}_{-0.20}$ & $-0.840^{+0.057}_{-0.084}$
        & $-0.46^{+0.34}_{-0.16}$ & $-0.43\pm0.21$ & $-0.838\pm0.054$ \\
        $w_a/w_\beta$
        & $<-0.536$ & $<-1.19$ & $-0.90^{+0.88}_{-0.23}$
        & $<-1.40$ & $-1.72\pm0.58$ & $-0.62^{+0.21}_{-0.19}$ \\
        $\beta$
        & $5.0^{+4.2}_{-6.8}$ & $1.8^{+1.2}_{-2.2}$ & $2.2^{+2.4}_{-5.2}$
        & --- & --- & --- \\
        \hline
        $H_0$
        & $>80.0$ & $64.8^{+1.7}_{-2.4}$ & $67.58\pm0.60$
        & $>79.2$ & $63.7^{+1.7}_{-2.1}$ & $67.54\pm0.60$ \\
        $\sigma_8$
        & --- & $0.790^{+0.016}_{-0.021}$ & $0.8125\pm0.0081$
        & --- & $0.782^{+0.015}_{-0.018}$ & $0.8126\pm0.0079$ \\
        $\Omega_m$
        & $0.328^{+0.022}_{-0.046}$ & $0.340^{+0.025}_{-0.021}$ & $0.3108\pm0.0058$
        & $0.354^{+0.041}_{-0.018}$ & $0.352\pm0.021$ & $0.3114\pm0.0057$ \\
        $S_8$
        & --- & $0.840^{+0.013}_{-0.011}$ & $0.8269\pm0.0083$
        & --- & $0.846\pm0.011$ & $0.8278\pm0.0077$ \\
        \hline
        $\Delta\chi^2_{\rm MAP}$
        & $-4.65$ & $-12.19$ & $-12.18$
        & $-4.69$ & $-12.53$ & $-10.22$ \\
        \hline\hline 
         & \multicolumn{3}{c}{\textbf{BA}} & \multicolumn{3}{c}{\textbf{JBP}} \\
        \cline{2-4} \cline{5-7}
        \textbf{Parameters} & \textbf{QB} & \textbf{QBC} & \textbf{QBCS} & \textbf{QB} & \textbf{QBC} & \textbf{QBCS} \\
        \hline
        $100\theta_{\rm MC}$
        & --- & $1.04079\pm0.00024$ & $1.04086\pm0.00024$
        & --- & $1.04091\pm0.00024$ & $1.04092\pm0.00024$ \\
        $\Omega_b h^2$
        & $0.0482^{+0.021}_{-0.0094}$ & $0.02221\pm0.00013$ & $0.02224\pm0.00012$
        & $0.048^{+0.022}_{-0.012}$ & $0.02227\pm0.00012$ & $0.02228\pm0.00012$ \\
        $\Omega_c h^2$
        & $0.221^{+0.077}_{-0.063}$ & $0.11959\pm0.00083$ & $0.11911\pm0.00079$
        & $0.180^{+0.060}_{-0.041}$ & $0.11872\pm0.00077$ & $0.11862\pm0.00077$ \\
        $n_s$
        & --- & $0.9645\pm0.0036$ & $0.9656\pm0.0036$
        & --- & $0.9666\pm0.0035$ & $0.9669\pm0.0035$ \\
        $\tau_{\rm reio}$
        & --- & $0.0529\pm0.0069$ & $0.0556\pm0.0070$
        & --- & $0.0581^{+0.0065}_{-0.0074}$ & $0.0588^{+0.0066}_{-0.0073}$ \\
        $\log(10^{10} A_s)$
        & --- & $3.039\pm0.013$ & $3.045\pm0.013$
        & --- & $3.050\pm0.013$ & $3.051\pm0.013$ \\
        \hline
        $w_0$
        & $-0.35\pm0.32$ & $-0.50\pm0.19$ & $-0.860\pm0.047$
        & $-0.64^{+0.31}_{-0.15}$ & $-0.652^{+0.23}_{-0.077}$ & $-0.811^{+0.072}_{-0.082}$ \\
        $w_a$
        & $-1.21\pm0.65$ & $-0.84^{+0.31}_{-0.26}$ & $-0.311^{+0.10}_{-0.093}$
        & $<-0.995$ & $<-1.69$ & $-1.15^{+0.49}_{-0.42}$ \\
        \hline
        $H_0$
        & $>79.1$ & $63.7\pm1.9$ & $67.55\pm0.60$
        & $>79.3$ & $66.5^{+1.0}_{-1.9}$ & $67.58\pm0.60$ \\
        $\sigma_8$
        & --- & $0.782\pm0.017$ & $0.8128\pm0.0079$
        & --- & $0.802^{+0.012}_{-0.018}$ & $0.8105\pm0.0078$ \\
        $\Omega_m$
        & $0.378\pm0.043$ & $0.352^{+0.020}_{-0.023}$ & $0.3113\pm0.0057$
        & $0.319^{+0.026}_{-0.015}$ & $0.321^{+0.018}_{-0.010}$ & $0.3100\pm0.0057$ \\
        $S_8$
        & --- & $0.846\pm0.012$ & $0.8278\pm0.0077$
        & --- & $0.8283^{+0.0097}_{-0.0078}$ & $0.8238\pm0.0075$ \\
        \hline
        $\Delta\chi^2_{\rm MAP}$
        & $-4.64$ & $-13.00$ & $-11.41$
        & $-3.29$ & $-8.57$ & $-7.40$ \\
        \hline\hline
    \end{tabular*}
    \caption{Constraints on the cosmological parameters for the GEoS, CPL, BA, and JBP parametrizations, across three data combinations (QB, QBC, and QBCS). Negative values of $\Delta\chi^2_{\rm MAP}$ indicate a goodness-of-fit improvement relative to the $\Lambda$CDM model. \vspace{2pt} \newline {\footnotesize \textit{Note.} $H_0$ is in units of $\mathrm{km\,s^{-1}\,Mpc^{-1}}$. All other parameters are dimensionless.}}
    \label{tab:combined_de_models}
\end{table*}

\begin{figure*}[t]
    \centering
    \includegraphics[width=0.8\textwidth]{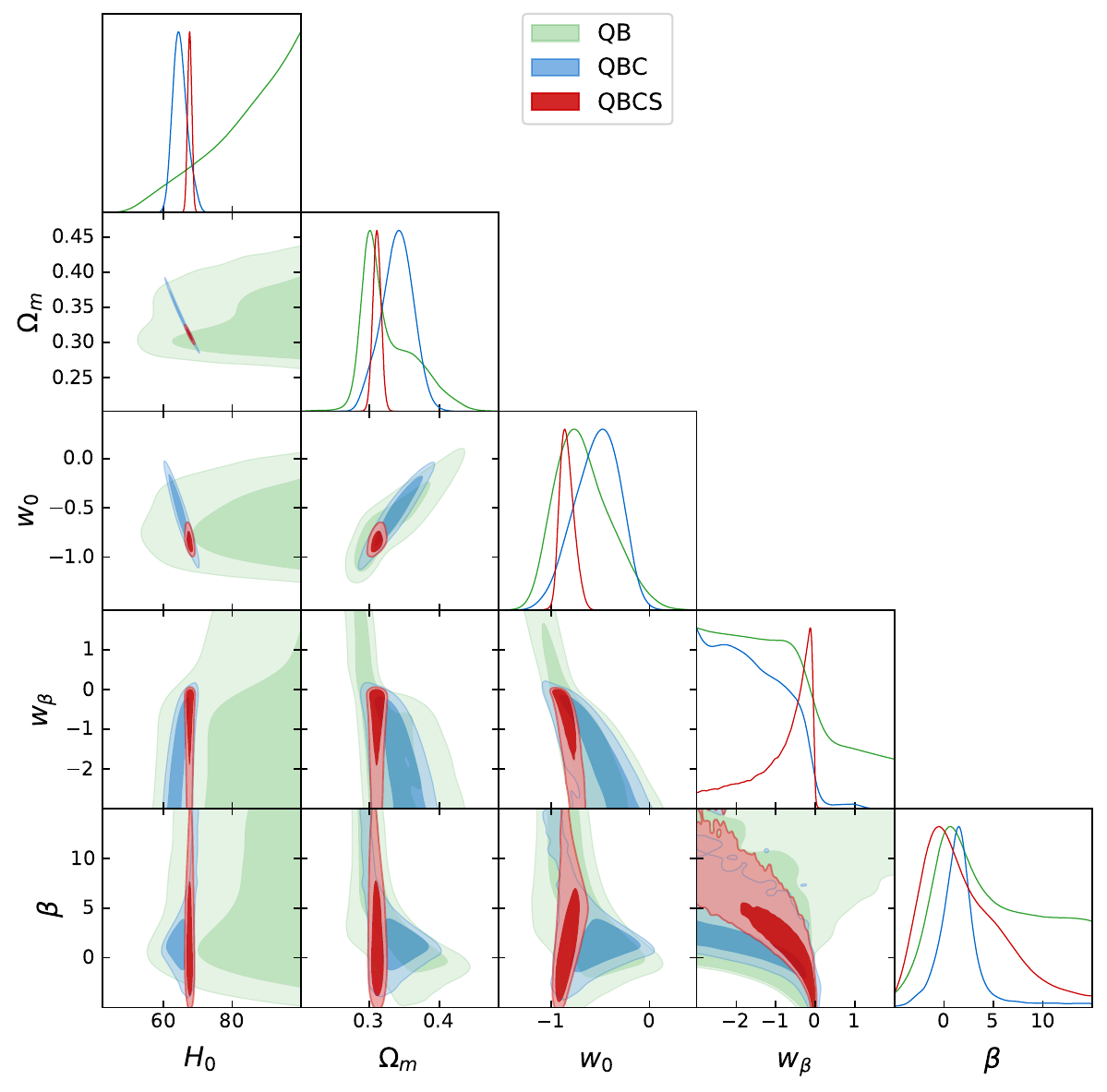}
    \caption{The one-dimensional posterior distributions and two-dimensional marginalized contours for the parameters in the GEoS parametrization, obtained from the QB, QBC, and QBCS datasets.}
    \label{ag1_plot}
\end{figure*}

\begin{figure*}[t]
    \centering
    \includegraphics[width=0.8\textwidth]{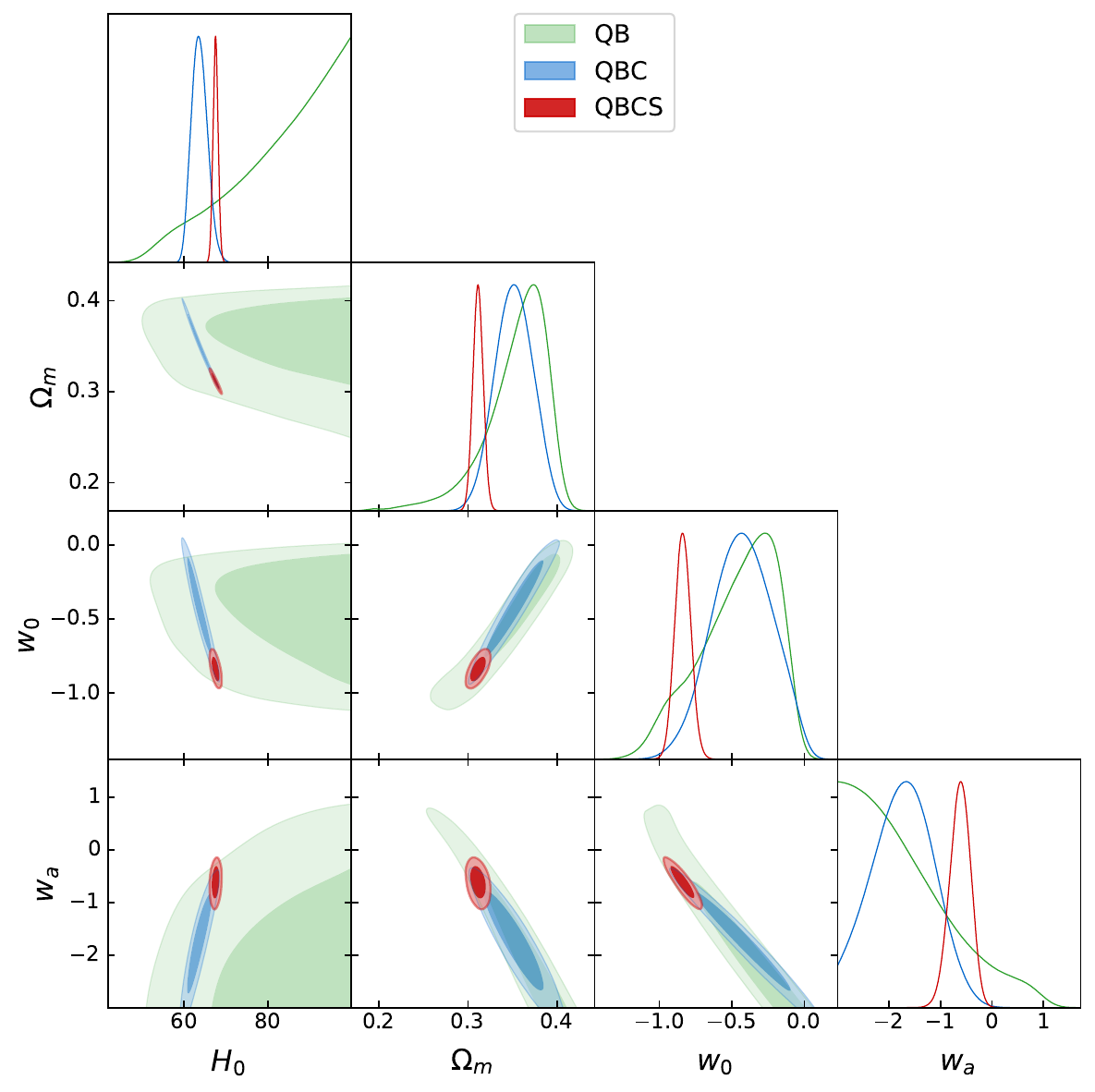}
    \caption{The one-dimensional posterior distributions and two-dimensional marginalized contours for the parameters in the CPL parametrization, obtained from the QB, QBC, and QBCS datasets.}
    \label{cpl_plot}
\end{figure*}

\begin{figure*}[t]
    \centering
    \includegraphics[width=0.8\textwidth]{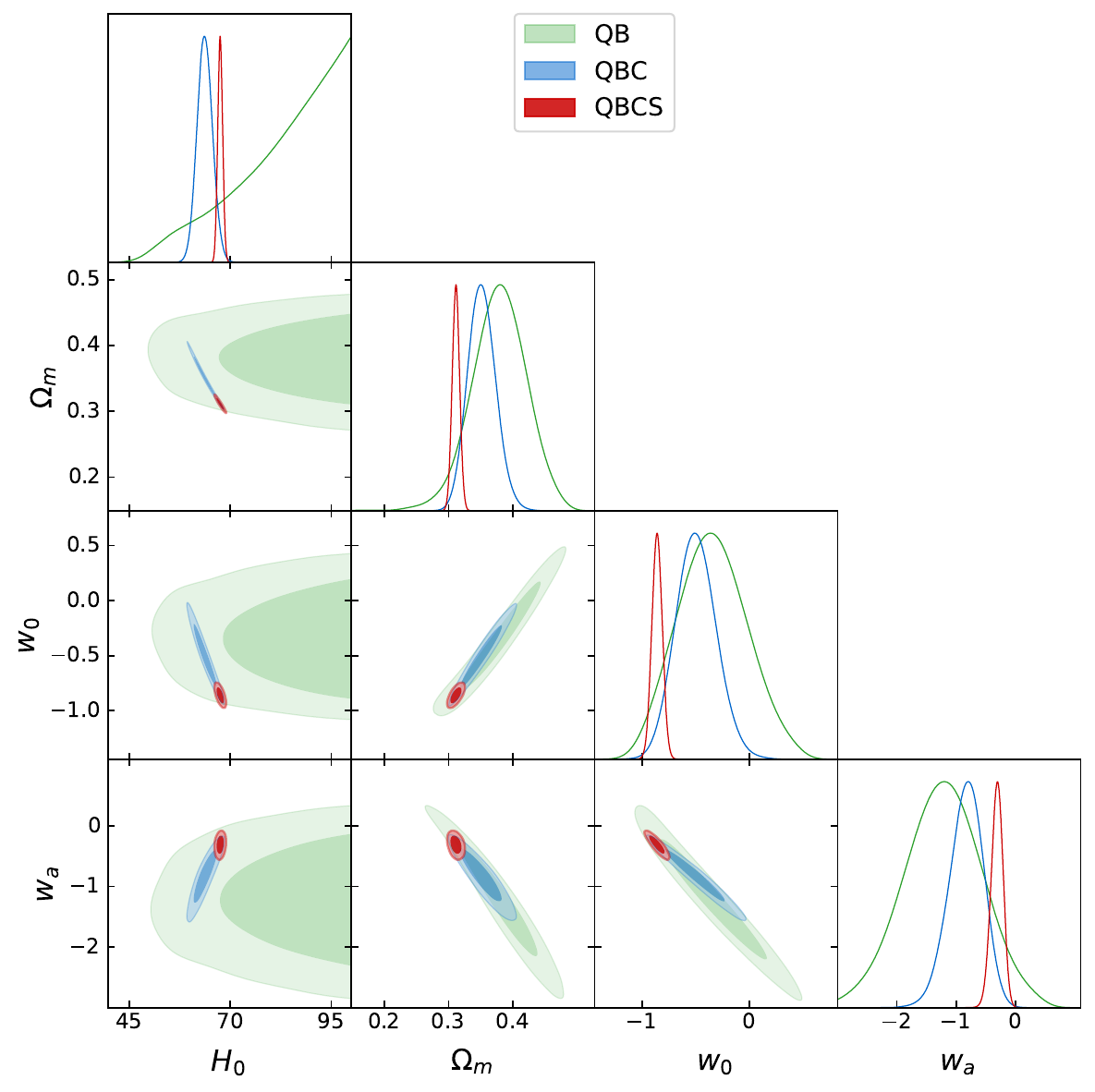}
    \caption{The same as Fig.~\ref{cpl_plot}, but for the BA parametrization.}
    \label{ba_plot}
\end{figure*}

\begin{figure*}[t]
    \centering
    \includegraphics[width=0.8\textwidth]{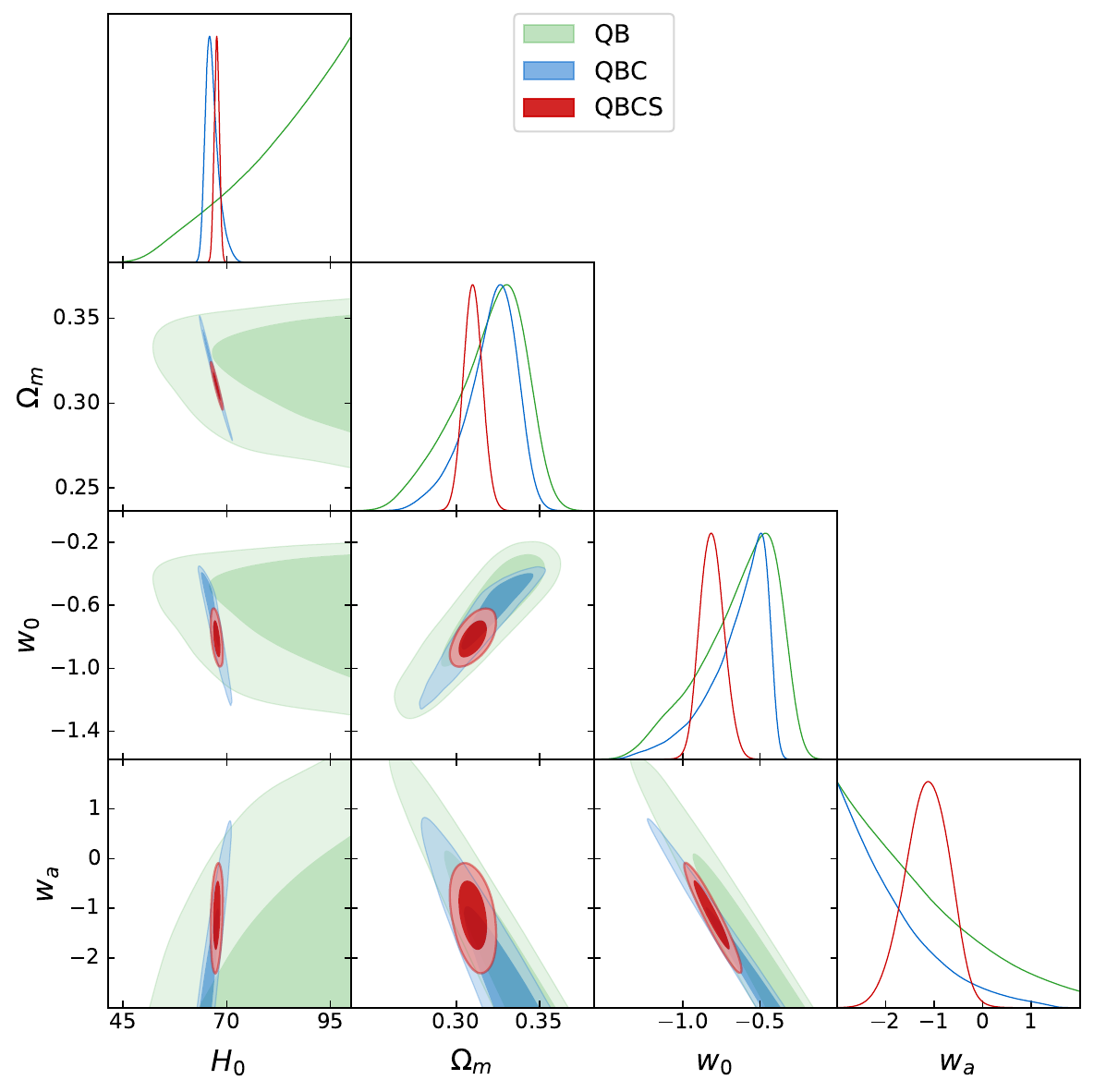}
    \caption{The same as Fig.~\ref{cpl_plot}, but for the JBP parametrization.}
    \label{jbp_plot}
\end{figure*}

\begin{figure*}[t]
    \centering
    \includegraphics[width=1.5\columnwidth]{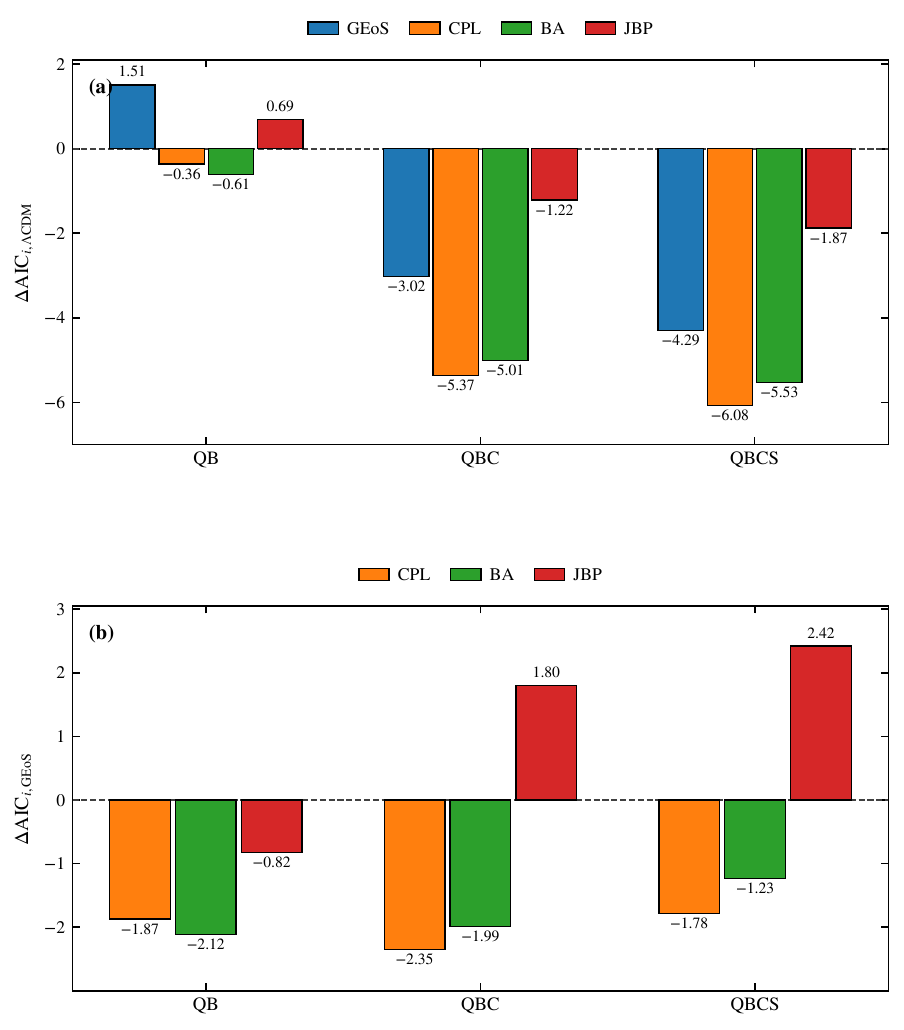}
    \caption{AIC differences among the dark-energy models for the QB, QBC,
    and QBCS data combinations. Panel (a) shows the AIC differences of
    the GEoS, CPL, BA, and JBP models relative to $\Lambda$CDM,
    $\Delta{\rm AIC}_{i,\Lambda{\rm CDM}}
    ={\rm AIC}_i-{\rm AIC}_{\Lambda{\rm CDM}}$.
    Panel (b) shows the AIC differences of the CPL, BA, and JBP models
    relative to GEoS, $\Delta{\rm AIC}_{i,{\rm GEoS}}
    ={\rm AIC}_i-{\rm AIC}_{\rm GEoS}$.
    The horizontal dashed line denotes $\Delta{\rm AIC}=0$. Negative values indicate a preference for model $i$ over the corresponding reference model,
    whereas positive values indicate a preference for the reference model.}
    \label{fig:AIC_compare}
\end{figure*}

\begin{figure*}[t]
    \centering
    \includegraphics[width=1.5\columnwidth]{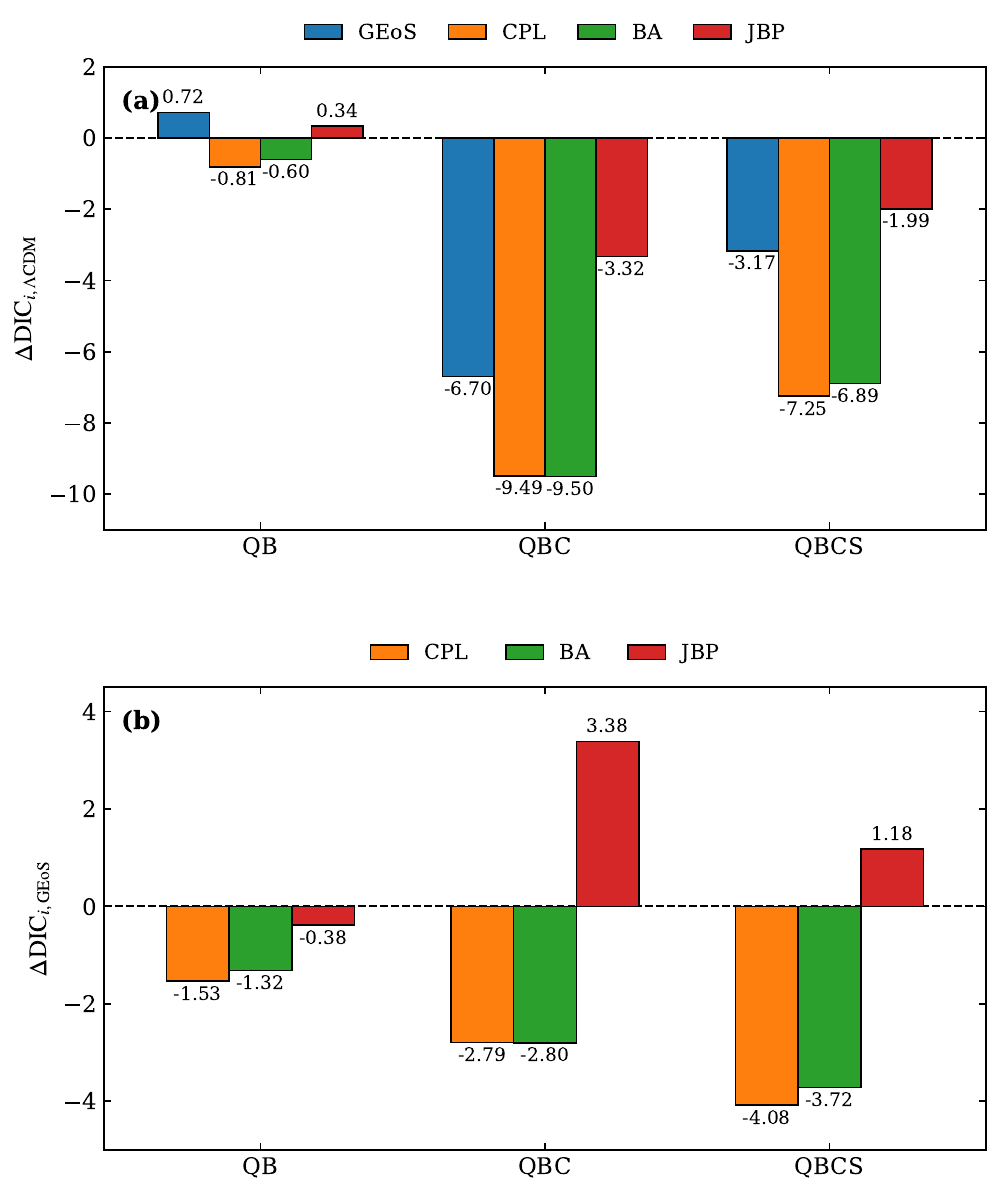}
    \caption{DIC differences among the dark-energy models for the QB, QBC,
    and QBCS data combinations. Panel (a) shows the DIC differences of
    the GEoS, CPL, BA, and JBP models relative to $\Lambda$CDM,
    $\Delta{\rm DIC}_{i,\Lambda{\rm CDM}}
    ={\rm DIC}_i-{\rm DIC}_{\Lambda{\rm CDM}}$.
    Panel (b) shows the DIC differences of the CPL, BA, and JBP models
    relative to GEoS,
    $\Delta{\rm DIC}_{i,{\rm GEoS}}
    ={\rm DIC}_i-{\rm DIC}_{\rm GEoS}$.}
    \label{fig:DICcompare}
\end{figure*}

\begin{figure*}[t]
    \centering
    \includegraphics[width=0.8\textwidth]{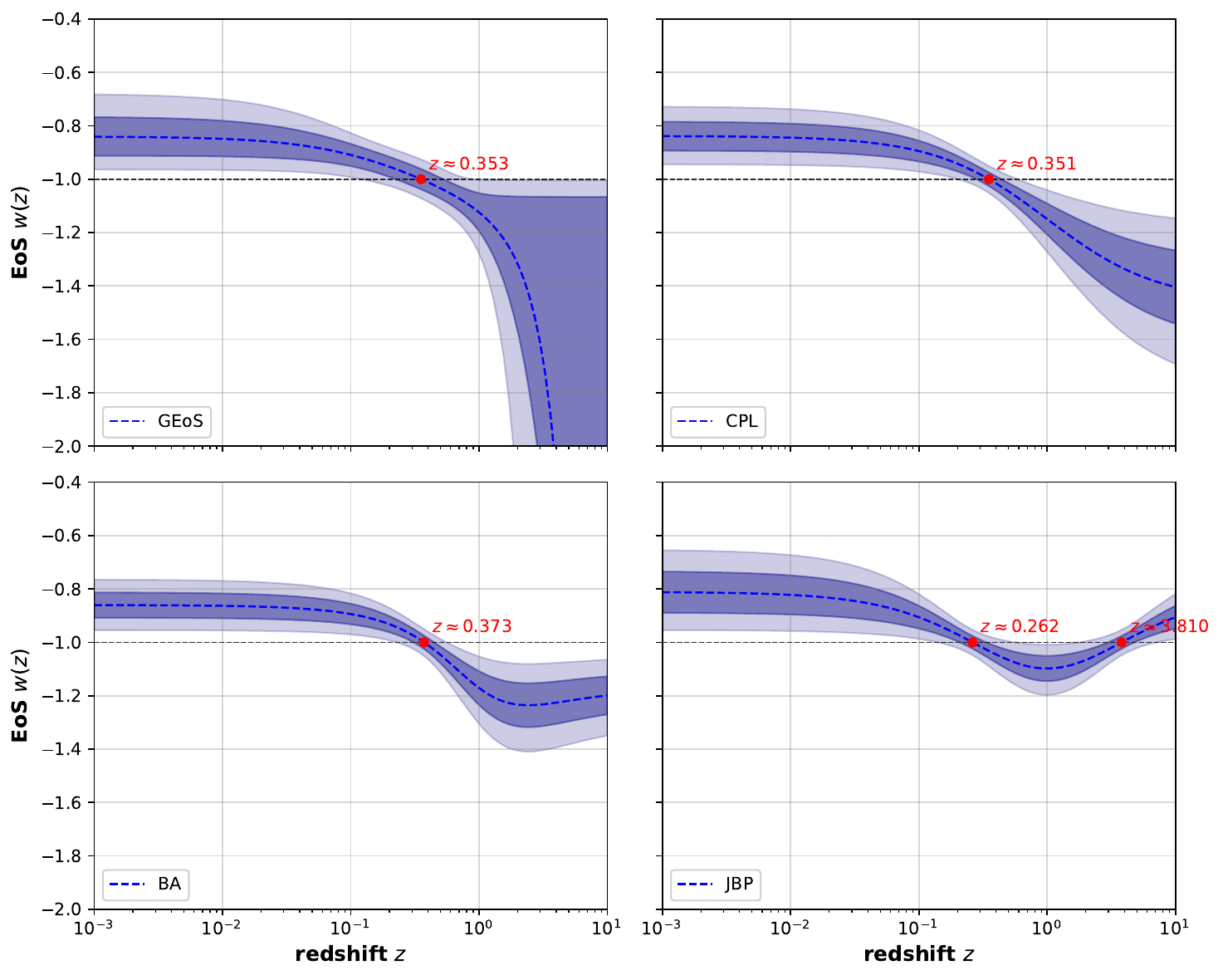}
    \caption{Redshift evolution of $w(z)$ over the redshift range $0 < z < 10$, obtained for the four models considered in this analysis.  Dashed lines represent the mean posterior values, while dark and light shaded contours denote the $1\sigma$ and $2\sigma$ C. L., respectively.}
    \label{wz_plot}
\end{figure*}

\section{Results And Discussion}
\label{sec:results}

In this section, we present the observational constraints on the four DDE models considered in this work using three data combinations: Quasars + DESI BAO (QB), Quasars + DESI BAO + CMB (QBC), and Quasars + DESI BAO + CMB + SNe Ia (QBCS). The marginalized constraints on the cosmological parameters for the four DDE models are summarized in Table~\ref{tab:combined_de_models}. The corresponding one-dimensional marginalized posterior distributions and two-dimensional marginalized confidence contours for the key cosmological parameters are shown in Fig.~\ref{ag1_plot} (GEoS), Fig.~\ref{cpl_plot} (CPL), Fig.~\ref{ba_plot} (BA), and Fig.~\ref{jbp_plot} (JBP), respectively. Using the joint data combinations, we assess the relative performance of the dark-energy models through the AIC and DIC difference. As shown in Fig.~\ref{fig:AIC_compare} and Fig.~\ref{fig:DICcompare}, we compare the GEoS, CPL, BA, and JBP models with the standard $\Lambda$CDM model, and then compare the CPL, BA, and JBP
parametrizations directly with GEoS. Fig.~\ref{wz_plot} shows the posterior mean evolution of $w(z)$ (dashed lines), together with the corresponding 68\% (dark shaded regions) and 95\% (light shaded regions) C.L. over the redshift range $0<z<10$, obtained from the full QBCS data combination.

\begin{table}
\centering
\renewcommand{\arraystretch}{1.4}
\begin{tabular}{lccc}
\hline
\diagbox{Datasets}{Parameters}
& $w_0$  & $w_\beta$  & $\beta$  \\
\hline
P-LBS  & $-0.842^{+0.060}_{-0.083}$ & $-0.94^{+0.94}_{-0.31}$ & $2.8^{+2.8}_{-5.9}$ \\
AW-LBS & $-0.838^{+0.064}_{-0.084}$ & $-1.04^{+1.0}_{-0.40}$ & $3.7^{+3.4}_{-5.9}$ \\
QBCS              & $-0.840^{+0.057}_{-0.084}$ & $-0.90^{+0.88}_{-0.23}$ & $2.2^{+2.4}_{-5.2}$ \\
\hline
\end{tabular}
\caption{Comparison of EoS parameter estimation for the GEoS model from different data combination. P-LBS and AW-LBS originate from two separate observational datasets in the previous works~\cite{Lee_2026}, while QBCS serves as the dataset used in this analysis.}
\label{tab:param_constraint_compare}
\end{table}

\subsection{GEoS parametrization}

Using the baseline QB data combination for the GEoS model, we obtain
$w_0=-0.67^{+0.21}_{-0.34}$. The posterior mildly shifts toward the
quintessence regime, although $w_0=-1$ remains within the 68\% credible
interval. The parameter $w_\beta$ is constrained only by an upper limit,
$w_\beta<-0.536$ at the 68\% C. L., placing $w_\beta=0$ outside
the corresponding credible region. In contrast, $\beta$ remains weakly
constrained, with $\beta=5.0^{+4.2}_{-6.8}$. Therefore, although the QB
data provide some indication of a departure from the constant-$w$ limit,
they do not tightly constrain the detailed evolution of the dark-energy EoS.

Compared with the QB data combination, adding CMB observations to form QBC substantially reduces the parameter degeneracies. The posterior of $w_0$ shifts further toward the
quintessence regime, yielding $w_0=-0.54^{+0.26}_{-0.20}$, with $w_0=-1$
lying outside the quoted 68\% credible interval. Meanwhile, the upper
limit on $w_\beta$ shifts toward more negative values,
$w_\beta<-1.19$. Adding CMB data also considerably tightens the
constraint on $\beta$, yielding $\beta=1.8^{+1.2}_{-2.2}$. The resulting interval encompasses the CPL limit $\beta=1$. Nevertheless, $w_\beta$
remains constrained only from above, preventing a precise determination
of the detailed redshift evolution of the dark-energy EoS.

When Pantheon+ SNe Ia observations are incorporated into the full QBCS
data combination, both $w_\beta$ and $\beta$ acquire two-sided
constraints. We obtain $w_0=-0.840^{+0.057}_{-0.084}$, with substantially
improved precision compared with the QB and QBC results. The parameter
$w_\beta$ is constrained to $w_\beta=-0.90^{+0.88}_{-0.23}$, with the
68\% credible interval extending to approximately $-0.02$. Thus,
$w_\beta=0$, corresponding to the constant-$w$ limit, lies just outside
the 68\% credible region. The remaining GEoS parameter is constrained to be $\beta=2.2^{+2.4}_{-5.2}$. Despite the improvement relative to the QB
constraint, its relatively broad posterior indicates that a wide range
of evolutionary behaviors remains compatible with the data.

We further compare the parameter constraints obtained from the full QBCS
data combination with two sets of results reported in
Ref.~\cite{Lee_2026}, as summarized in
Table~\ref{tab:param_constraint_compare},  where ``P'', ``AW'', ``L'', ``B'', and ``S'' denote the Planck 2018 primary CMB, the Planck-independent primary CMB (ACT DR6 + WMAP), the combined CMB lensing (ACT DR6 + Planck PR4 NPIPE), the DESI DR2 BAO, and the Pantheon+ SNIa datasets, respectively. The three sets of constraints
are broadly consistent. For the present-day EoS parameter, we obtain
$w_0=-0.840^{+0.057}_{-0.084}$, whose central value lies between those
reported for the two literature data combinations. Our analysis also
yields a narrower marginalized constraint on $w_\beta$, particularly
toward negative values. Its 68\% upper bound reaches approximately
$-0.02$, placing $w_\beta=0$ just outside the corresponding credible
interval. Compared with the $w_\beta$ constraints obtained from the
AW-LBS data combination, this indicates an improved constraint on
deviations from the constant-$w$ limit. The parameter $\beta=2.2^{+2.4}_{-5.2}$, however, remains relatively
weakly constrained, primarily due to its strong nonlinear degeneracy
with $w_\beta$, as evident from the two-dimensional marginalized
posterior contours.
Nevertheless, compared with the $\beta$ constraints obtained from the
P-LBS and AW-LBS data combinations in
Table~\ref{tab:param_constraint_compare}, the QBCS analysis yields a
narrower marginalized interval and therefore reduces the range of
$\beta$ values allowed by the data.

\subsection{Other DDE parametrizations}

\textbf{CPL parametrization}: For the QB data combination, we obtain
$w_0=-0.46^{+0.34}_{-0.16}$ at the 68\% C. L. The posterior
of $w_0$ lies in the quintessence regime, with $w_0=-1$ outside the
quoted 68\% credible interval. For the dynamical parameter $w_a$, the
QB data yield a one-sided upper limit of $w_a<-1.40$ at the 68\%
C. L. The upper limit lies below $w_a=0$, indicating a
preference for negative values of $w_a$. Nevertheless, the absence of
a two-sided constraint on $w_a$ indicates that the QB data alone do
not tightly determine the dynamical evolution of the CPL EoS.

The inclusion of CMB observations substantially improves the
constraints. For the QBC data combination, we obtain
$w_0=-0.43\pm0.21$, with the posterior remaining in the quintessence
regime and $w_0=-1$ lying outside the 68\% credible interval. The
dynamical parameter is constrained to $w_a=-1.72\pm0.58$, corresponding
to a clear preference for negative $w_a$. Within the CPL
parametrization, the negative value of $w_a$, together with $w_0>-1$,
implies a decreasing EoS with increasing redshift, with the central
reconstructed evolution crossing into the phantom regime at higher
redshifts.

Adding the Pantheon+ SNe Ia data substantially reduces the allowed
parameter space and tightens the constraints on the dark-energy
parameters. The full QBCS analysis yields
$w_0=-0.838\pm0.054$ and
$w_a=-0.62^{+0.21}_{-0.19}$, in excellent agreement with the results
reported by the DESI Collaboration,
$w_0=-0.838\pm0.055$ and
$w_a=-0.62^{+0.22}_{-0.19}$~\cite{Abdul_Karim_2025}. Compared with the QBC result, the present-day EoS parameter $w_0$
shifts closer to the cosmological-constant value, although its
68\% credible interval remains entirely within the quintessence
regime. Meanwhile, $w_a$ shifts toward a less negative value and is
constrained on both sides, with $w_a=0$ lying outside the quoted
68\% credible interval. These results indicate that the inclusion of
Pantheon+ substantially improves the determination of the CPL
parameters while retaining a preference for an evolving dark-energy
equation of state.

\textbf{BA parametrization}: For the QB data combination, we obtain
$w_0=-0.35\pm0.32$ at the 68\% C. L. The marginalized
posterior favors $w_0>-1$, although the constraint remains relatively
weak. The dynamical parameter is constrained to
$w_a=-1.21\pm0.65$, indicating a preference for negative values of
$w_a$. Under a one-dimensional Gaussian approximation, the deviations
from $w_0=-1$ and $w_a=0$ correspond to approximately $2.0\sigma$ and
$1.9\sigma$, respectively. These modest deviations indicate that the
QB data alone do not provide a strong indication of an evolving
dark-energy EoS.

For the QBC data combination, the constraints are substantially
improved, yielding $w_0=-0.50\pm0.19$. The dynamical parameter is constrained to
$w_a=-0.84^{+0.31}_{-0.26}$, with $w_a=0$ lying outside the quoted
68\% credible interval. These results indicate that the inclusion of
CMB observations substantially improves the determination of both
dark-energy parameters.

With the addition of Pantheon+ SNe Ia observations, the full QBCS
data combination yields substantially tighter constraints,
$w_0=-0.86\pm0.047$ and
$w_a=-0.311^{+0.100}_{-0.093}$ at the 68\% C. L. These
constraints are in excellent agreement with the results reported in
Ref.~\cite{2026PDU....5202254L},
$w_0=-0.856\pm0.048$ and $w_a=-0.32\pm0.09$.
The marginalized constraint on $w_0$ corresponds to an approximately
$3.0\sigma$ deviation from $w_0=-1$ under a one-dimensional Gaussian
approximation. Similarly, the constraint on $w_a$ places $w_a=0$
outside the 68\% credible interval, with its central value lying
approximately $3.1\sigma$ from zero when estimated using the upper
uncertainty. Both marginalized constraints therefore favor
$w_0>-1$ and $w_a<0$, indicating a preference for an evolving
dark-energy EoS within this parametrization.

\textbf{JBP parametrization}: For the QB data combination, we obtain $w_0=-0.64^{+0.31}_{-0.15}$ at the 68\% C. L. The
marginalized posterior favors $w_0>-1$, with $w_0=-1$ lying outside the quoted 68\% credible interval. Meanwhile, the dynamical parameter $w_a$ remains unconstrained from below, yielding only a one-sided upper limit of $w_a<-0.995$ at the 68\% C. L. These results indicate that the QB data alone provide only limited constraints on the dynamical evolution of the JBP EoS.

For the QBC data combination, we obtain
$w_0=-0.652^{+0.23}_{-0.077}$. The marginalized posterior remains
asymmetric, with a substantially reduced lower uncertainty, while
$w_0=-1$ remains outside the quoted 68\% credible interval. The
parameter $w_a$ remains constrained only from above, with
$w_a<-1.69$ at the 68\% C. L. Therefore, although the
inclusion of CMB observations improves the constraint on $w_0$, the
QBC data still do not provide a two-sided constraint on $w_a$.

With the addition of Pantheon+ SNe Ia observations, the full QBCS
data combination yields $w_0=-0.811^{+0.072}_{-0.082}$ and
$w_a=-1.15^{+0.49}_{-0.42}$ at the 68\% C. L. Compared
with the QB and QBC results, the addition of SNe Ia data substantially
improves the determination of both JBP parameters and provides a
two-sided constraint on $w_a$. The marginalized posterior of $w_0$
lies in the quintessence regime, with $w_0=-1$ outside the quoted
68\% credible interval. Meanwhile, the constraint on $w_a$ favors
negative values, with $w_a=0$ also lying outside the quoted 68\%
credible interval. The marginalized posterior favors negative values of $w_a$,
suggesting a redshift evolution of the EoS within the JBP
parametrization.

\subsection{Statistical analysis}

We compare the performance of the DDE models with the standard
$\Lambda$CDM framework, with the corresponding fitting statistics
summarized in Table~\ref{tab:combined_de_models}. We define
$\Delta\chi^2_{\rm MAP}=\chi^2_{\rm MAP,DDE}-\chi^2_{\rm MAP,\Lambda{\rm CDM}}$.
The resulting $\Delta\chi^2_{\rm MAP}$ values are negative for all
DDE models across the data combinations considered, indicating that
the DDE models provide improved goodness of fit relative to
$\Lambda$CDM. For the full QBCS data combination, the GEoS model yields
the largest improvement, with $\Delta\chi^2_{\rm MAP}=-12.18$. To account for the effective model complexity in addition to the
quality of fit, we further compare the models using the AIC and DIC. We take $\Lambda$CDM as the reference model and compare the
four DDE parametrizations, as shown in Fig.~\ref{fig:AIC_compare}(a) and Fig.~\ref{fig:DICcompare}(a). 

For the QB data combination, the AIC differences between the four DDE models and $\Lambda$CDM are small. The corresponding
$\Delta{\rm AIC}$ values for GEoS, CPL, BA, and JBP are
$1.51$, $-0.36$, $-0.61$, and $0.69$, respectively.
Negative values indicate a preference for DDE, while positive values indicate a preference for $\Lambda$CDM.
The absolute values of these differences are all smaller than $2$,
indicating only weak discrimination between the DDE models and
$\Lambda$CDM according to the adopted AIC criterion.
The corresponding DIC differences are likewise small, with
$\Delta{\rm DIC}=0.72$, $-0.81$, $-0.60$, and $0.34$
for GEoS, CPL, BA, and JBP, respectively. Their magnitudes are all
well below the reference value of $|\Delta{\rm DIC}|=5$, indicating
only weak discrimination between the DDE models and $\Lambda$CDM
according to the adopted DIC criterion. After including the CMB data, the AIC differences of GEoS, CPL, BA,
and JBP relative to $\Lambda$CDM become
$-3.02$, $-5.37$, $-5.01$, and $-1.22$, respectively.
All four values are negative, indicating a preference for the DDE models. The absolute difference for GEoS
exceeds the weak-discrimination range, while those for CPL and BA
reach the moderate-discrimination range. In contrast, the AIC
difference for JBP remains within the weak-discrimination range. The corresponding DIC differences are
$-6.70$, $-9.49$, $-9.50$, and $-3.32$, respectively.
Thus, all four DDE models also yield lower DIC values than
$\Lambda$CDM. The differences for GEoS, CPL, and BA exceed the
reference value of $|\Delta{\rm DIC}|=5$ associated with moderate
discrimination, with those for CPL and BA approaching the reference
value of $|\Delta{\rm DIC}|=10$ associated with strong
discrimination. In contrast, the difference for JBP remains below
the moderate reference value. With the further inclusion of the Pantheon+ SNe Ia data, the AIC differences remain negative for all four DDE parametrizations.
The corresponding $\Delta{\rm AIC}$ values for GEoS, CPL, BA, and
JBP are $-4.29$, $-6.08$, $-5.53$, and $-1.87$, respectively.
The absolute differences for GEoS, CPL, and BA lie within the
moderate-discrimination range, whereas the difference for JBP remains
within the weak-discrimination range. The DIC differences also remain negative, with $\Delta{\rm DIC}=-3.17$, $-7.25$, $-6.89$, and $-1.99$
for GEoS, CPL, BA, and JBP, respectively. The differences for CPL
and BA remain above the moderate reference value, whereas those for
GEoS and JBP fall below it. Taken together, the AIC and DIC results
show that the degree of discrimination between the DDE models and
$\Lambda$CDM depends appreciably on the adopted data combination.

To further examine the role of the additional flexibility introduced
by GEoS, we next take GEoS as the reference model and compare it
directly with the CPL, BA, and JBP parametrizations, as shown in
Fig.~\ref{fig:AIC_compare}(b) and Fig.~\ref{fig:DICcompare}(b). For the QB data combination, the $\Delta{\rm AIC}$ values of CPL, BA,
and JBP relative to GEoS are $-1.87$, $-2.12$, and $-0.82$,
respectively. All three values are negative, indicating that these
two-parameter DDE parametrizations yield lower AIC values than GEoS.
Among them, the absolute AIC difference for BA exceeds the
weak-discrimination range, while those for CPL and JBP remain within
or close to the weak-discrimination range. The corresponding $\Delta{\rm DIC}$ values are $-1.53$, $-1.32$, and $-0.38$ for CPL, BA, and JBP, respectively.
Their absolute values remain well below the moderate reference value
of $|\Delta{\rm DIC}|=5$, indicating only weak discrimination
relative to GEoS. After including the CMB data, the corresponding AIC differences
become $\Delta{\rm AIC}=-2.35$, $-1.99$, and $+1.80$
for CPL, BA, and JBP, respectively. Thus, CPL and BA yield lower AIC
values than GEoS, whereas GEoS yields a lower AIC than JBP.
The corresponding DIC differences are $\Delta{\rm DIC}=-2.79$, $-2.80$, and $+3.38$, respectively, showing the same qualitative ordering. For the full QBCS data combination, the AIC differences are $\Delta{\rm AIC}=-1.78$, $-1.23$, and $+2.42$ for CPL, BA, and JBP, respectively, while the corresponding DIC
differences are $\Delta{\rm DIC}=-4.08$, $-3.72$, and $+1.18$.
This indicates that the CPL and BA models have negative
$\Delta \mathrm{DIC}$ and $\Delta \mathrm{AIC}$ values, implying a preference for these models; however, the absolute magnitudes of these differences remain below the moderate reference. For the JBP model, both $\Delta \mathrm{DIC}$ and $\Delta \mathrm{AIC}$ are positive, indicating a preference for the GEoS model. Its $\Delta \mathrm{DIC}$ is less than 5, suggesting below the moderate value, whereas its $\Delta \mathrm{AIC}$ exceeds 2, going beyond the weak-evidence range. Overall, the AIC and DIC comparisons show a broadly consistent
pattern. CPL and BA generally yield lower information-criterion
values than GEoS, whereas GEoS is favored over JBP after the inclusion
of CMB data. Nevertheless, the relatively small differences among
the DDE parametrizations indicate that the current observations do
not provide a clear statistical discrimination among these dynamical
dark-energy scenarios.

To illustrate the evolutionary behavior of the four DDE
parametrizations, we reconstruct their equations of state $w(z)$ over
the redshift range $0<z<10$, as shown in Fig.~\ref{wz_plot}. The blue
dashed curves represent the central reconstructed evolution, while the
dark and light shaded regions denote the corresponding 68\% and 95\%
C. L., respectively. At $z>1$, the GEoS parametrization
exhibits the broadest uncertainty band, reflecting both the limited
constraining power of the current data on the high-redshift EoS and
the additional parameter degeneracies present in this parametrization. The central reconstructed trajectories of the GEoS, CPL, and BA models
exhibit a qualitatively similar evolutionary trend, evolving from the
phantom regime ($w<-1$) at higher redshifts toward the quintessence
regime ($w>-1$) at late times. Their central trajectories cross the
cosmological-constant boundary $w=-1$ at approximately
$z=0.353$, $0.351$, and $0.373$, respectively. In contrast, the JBP
parametrization exhibits a non-monotonic evolution. Moving from high to
low redshift, its central trajectory crosses from the quintessence
regime into the phantom regime at $z\simeq3.810$, reaches a minimum at
$z=1$, and subsequently turns upward, recrossing $w=-1$ at
$z\simeq0.262$ before entering the quintessence regime at late times.

\section{Conclusions}
\label{sec:conclusions}

Recent DESI DR2 results, when combined with complementary cosmological probes, have strengthened indications of a time-varying dark-energy EoS. These results are commonly characterized within the CPL parametrization, motivating an examination of whether similar evolutionary behavior persists in alternative DDE parametrizations. In this work, we therefore combine multiple cosmological probes to constrain and compare several phenomenological models of dynamical dark energy. Our primary focus is the three-parameter GEoS model, which encompasses several commonly studied DDE parametrizations as special or limiting cases. Specifically, it reduces to the CPL parametrization for $\beta=1$, approaches the logarithmic form in the limit $\beta\to0$, and corresponds to a linear-redshift parametrization for $\beta=-1$. For other values of $\beta$, the GEoS model allows for a broader range of dark-energy evolutionary behaviors. For comparison, we also consider the two-parameter CPL, BA, and JBP parametrizations. These models are constrained using combinations of quasar, DESI DR2 BAO, CMB, and Pantheon+ SNe Ia observations, with the marginalized parameter constraints summarized in Table~\ref{tab:combined_de_models}.

For the GEoS model, the full QBCS data combination yields $w_0=-0.840^{+0.057}_{-0.084}$, $w_\beta=-0.90^{+0.88}_{-0.23}$, and $\beta=2.2^{+2.4}_{-5.2}$ at the 68\% C. L. The marginalized posterior of $w_0$ favors the quintessence regime, while the 68\% credible interval of $w_\beta$ extends to approximately $-0.02$, placing $w_\beta=0$ just outside the quoted interval. The parameter $\beta$, however, remains relatively weakly constrained, primarily because of its nonlinear degeneracy with $w_\beta$. The QBCS constraints therefore favor negative $w_\beta$, while still allowing a relatively broad range of evolutionary behaviors within the GEoS parametrization. For the CPL, BA, and JBP parametrizations, the QBCS data yield $w_0=-0.838\pm0.054$, $-0.860\pm0.047$, and $-0.811^{+0.072}_{-0.082}$, respectively. These three parametrizations therefore provide broadly consistent constraints on the present-day dark-energy EoS, with their marginalized posteriors favoring $w_0>-1$. The corresponding dynamical parameters are $w_a=-0.62^{+0.21}_{-0.19}$ for CPL, $w_a=-0.311^{+0.100}_{-0.093}$ for BA, and $w_a=-1.15^{+0.49}_{-0.42}$ for JBP. In all three cases, the marginalized constraints favor negative values of $w_a$, with $w_a=0$ lying outside the quoted 68\% C. L.

For the full QBCS data combination, both the AIC and DIC yield
lower values for all four DDE parametrizations than for
$\Lambda$CDM, with the largest reduction found for
the CPL and BA parametrizations. Among the DDE parametrizations,
the two criteria show a broadly consistent pattern: CPL and BA
yield lower information-criterion values than GEoS, whereas GEoS
yields lower values than JBP. However, the differences among the
DDE parametrizations are not large enough to provide a clear discrimination among these models, indicating that the additional flexibility of GEoS does not necessarily translate into
a more favorable statistical performance once model complexity is
taken into account. The reconstructed equations of state further illustrate the different evolutionary behaviors allowed by these parametrizations. The central reconstructed trajectories of the GEoS, CPL, and BA models exhibit a qualitatively similar evolution, transitioning from the phantom regime ($w<-1$) at higher redshifts to the quintessence regime ($w>-1$) at low redshifts and crossing the cosmological-constant boundary once. In contrast, the central JBP reconstruction exhibits a nonmonotonic evolution, crossing $w=-1$ twice within the redshift range considered. At $z>1$, the GEoS model also exhibits the broadest credible region among the models considered, reflecting both the limited sensitivity of the current data to the high-redshift EoS and the greater flexibility of the GEoS parametrization, particularly the nonlinear degeneracy between $w_\beta$ and $\beta$.

Future DESI data releases, together with observations from the recently
launched Nancy Grace Roman Space Telescope, are expected to provide
substantially improved observational leverage for studies of dynamical
dark energy. In particular, Roman's High-Latitude Time-Domain Survey
will use Type Ia supernovae to provide precise measurements of the
cosmic expansion history, while its wide-area observations will offer
complementary cosmological information from large-scale structure and
weak gravitational lensing~\cite{Eifler2021Roman,Wang2023Roman}. Combined with larger and better-calibrated high-redshift quasar samples,
these observations will improve both the statistical precision and
redshift coverage of future multi-probe analyses. Such improvements
will be particularly important for testing the additional evolutionary
freedom introduced by the GEoS parametrization. A key objective will be
to determine whether the broad posterior of $\beta$ and its nonlinear
degeneracy with $w_\beta$ can be substantially reduced by improved
observational constraints. The increased statistical precision and
redshift coverage will also provide tighter constraints on the
high-redshift behavior of $w(z)$, where the GEoS model currently
exhibits substantial uncertainty. Future multi-probe observations will
therefore provide more stringent tests of the dynamical dark-energy
scenarios considered in this work and their possible departure from
$\Lambda$CDM.

\section{acknowledgments}
\label{sec:acknowledgments}

This work is supported by National Natural Science Foundation of China (Nos. 12675058; 12433001); Beijing Natural Science Foundation (No. 1242021).

\section{Data availability statements} 
The data underlying this article will be shared on reasonable request to the corresponding author.

\bibliographystyle{apsrev4-2} 

\bibliography{reference}

\end{document}